\documentclass{emulateapj}
\usepackage{graphicx}
\usepackage{color}

\newcommand{\msun}{M_{\odot}}

\newcommand{\msunyr}{M_\odot~{\rm yr}^{-1}}

\begin{document}


\title{The Final Fates of Accreting Supermassive Stars}


\author{Hideyuki Umeda\altaffilmark{1}, Takashi Hosokawa\altaffilmark{2,3}, 
Kazuyuki Omukai\altaffilmark{4}, Naoki Yoshida\altaffilmark{3,5}}

\altaffiltext{1}{Department of Astronomy, The University of Tokyo, Tokyo 113-0033, Japan}
\altaffiltext{2}{Department of Physics, Kyoto University, Kyoto 606-8502, Japan}
\altaffiltext{3}{Department of Physics and Research Center for the Early Universe, 
The University of Tokyo, Tokyo 113-0033, Japan}
\altaffiltext{4}{Astronomical Institute, Tohoku University, Sendai 980-8578, Japan}
\altaffiltext{5}{Kavli Institute for the Physics and Mathematics
of the Universe, University of Tokyo, Kashiwa, Chiba 277-8583, Japan}

%



\begin{abstract}
The formation of supermassive stars (SMSs) via rapid mass accretion 
and their direct collapse into black holes (BHs) is a promising pathway for 
sowing seeds of supermassive BHs in the early universe. 
We calculate the evolution of rapidly accreting SMSs 
by solving the stellar structure equations 
including nuclear burning as well as general relativistic (GR) effects 
up to the onset of the collapse. 
We find that such SMSs have less concentrated structure 
than fully-convective counterpart, which is often postulated 
for non-accreting ones. 
This effect stabilizes the stars against GR instability even 
above the classical upper mass limit $\gtrsim 10^5~\msun$ derived 
for the fully-convective stars.
The accreting SMS begins to collapse at the higher mass with the
higher accretion rate. 
The collapse occurs when the nuclear fuel is exhausted
only for cases with $\dot M \lesssim 0.1~\msunyr$.
With $\dot{M} \simeq 0.3 - 1~\msunyr$, the star becomes GR-unstable
during the helium-burning stage at $M \simeq 2 - 3.5~\times 10^5~\msun$. 
In an extreme case with $10~\msunyr$,
the star does not collapse until the mass reaches $\simeq 8.0\times 10^5~\msun$,
where it is still in the hydrogen-burning stage.
We expect that BHs with roughly the same mass 
will be left behind after the collapse in all the cases.
\end{abstract}

\keywords{cosmology: theory -- early universe -- galaxies: formation 
-- stars: formation -- accretion -- general relativity}


\section{Introduction} \label{sec:intro}

Recent discovery of luminous quasars at $z > 6$ suggests the
existence of black holes with mass exceeding $10^9~\msun$
when the age of the Universe was less than one billon years \citep[e.g.,][]{Mortlock11,Wu15}.
The formation process of such supermassive black holes (SMBHs)
is largely unknown, and poses a serious challenge to the theory of
structure formation.

The so-called direct collapse scenario \citep{BL03} 
provides an attractive pathway for the early BH formation 
via a peculiar mode of the primordial star formation.
This scenario supposes the gravitational collapse of a primordial gas cloud
forming a supermassive star with a mass of $\ga 10^5 \msun$.
Specifically, the formation process proceeds in a two-step manner;
a very small ($\ll M_{\sun}$) protostar first forms at the densest part of
the cloud, and then grows in mass by
rapid accretion from the surrounding envelope 
\citep[e.g.,][]{Latif13,Inayoshi14,Becerra15}.
The accretion rate onto the protostar is expected to be
as large as $\sim 0.1 - 1~\msunyr$, 
with which the stellar mass will exceed $\ga 10^5~\msun$ in
less than the stellar lifetime.
Such a supermassive star, if successfully formed, eventually collapses into a BH
either by the core-collapse or by the general relativistic (GR) instability 
\citep[e.g.,][]{Iben63,Chandra64}.
Given the massive remnant BH as a seed, a SMBH with more 
than $10^9~\msun$ can be assembled through further growth 
by accretion and/or mergers in less than one billion years \citep[e.g.,][]{DiMatteo12}.  


The size and the internal structure of an accreting protostar \citep[e.g.,][]{OP03,Ohkubo09} 
and its resulting radiative feedback effect, which potentially 
terminates the mass accretion \citep[e.g.,][]{McKee08, Hosokawa11, Hosokawa16, Stacy16},
are key ingredients to set the final stellar mass \citep[e.g.,][]{Hirano15}.
Recent calculations show that the protostar 
structure is substantially modified if the accretion rate
exceeds a few $\times 10^{-2}~\msunyr$; 
the star inflates to have an extended radius of $10-100$~AU 
\citep[e.g.,][]{Hosokawa12,Hosokawa13,Schleicher13}.
Since such a fluffy protostar has low effective temperature and hence
very low UV luminosity, the radiative feedback does not disturb 
the accretion flow at least until the stellar mass exceeds $\sim 10^5~\msun$
as long as continuous and efficient gas accretion ensues.

In this {\it Letter}, we extend our stellar evolution calculations 
until the SMSs begin to collapse. 
Our calculations determine the final mass of the accreting SMSs
by incorporating the realistic conditions 
within the context of the direct collapse scenario.
Previous studies show that the GR instability causes
the collapse of a non-rotating SMS with $\gtrsim$ a few 
$\times~10^5~\msun$ 
but without considering its evolution under rapid mass accretion
\citep[e.g.,][]{Osaki66,Unno71,Fricke73,Fuller86}.
We show that gas accretion significantly alters the stellar structure.
With a very large accretion rate, a star continues to grow in mass
without collapsing by the GR instability 
until the stellar mass reaches 
$\sim 10^6~\msun$. Our calculations show the final fate
of such a SMS.


\section{Method and Model} \label{sec:method}

\begin{table}[htb]
 \begin{center}
  \caption{Final stellar mass and composition of the inner core}
 \begin{tabular}{lcccc}
   \hline
   \hline
 $\dot{M} [M_\odot$/yr] & 0.1 & 0.3& 1.0 & 10 \\
\hline
 $M_{\rm f}  [M_\odot]$  & 1.2 $\times 10^5$ & 1.9 $\times 10^5$ & 3.5 $\times 10^5$ & 8.0 $\times 10^5$ \\ 
 $Y$ (or $X$)  & 0.00 & 0.99 & 1.00  & (0.51)  \\ 

   \hline
 \end{tabular}
\label{tb:t1}
 \end{center}
\end{table}

We solve the stellar evolution equations using a 
modified code developed in \citet{Umeda09}, incorporating 
the GR effects in the post-Newtonian approximation
\citep{Fuller86}.
We adopt a primordial abundance pattern for 
an initial model with hydrogen and helium mass fractions 
$X=0.753$ and $Y=0.247$, 
and the same nuclear reaction networks 
as in \citet{Umeda02,Umeda05} and \citet{Umeda09}. 
A major difference of the code from that of \citet{Hosokawa12, Hosokawa13} 
is that no entropy generated at the accretion shock 
is assumed to be incorporated into the stellar interior.
This corresponds to $\eta = 0$ models
in \citet{Hosokawa13}, where $\eta$ is the fraction of the
accretion energy deposited in the stellar interior.
However, varying this setting only affects
the early evolution for the total stellar mass $M \lesssim 100~\msun$.


We follow the evolution of an accreting protostar with four 
different rates of $\dot M$ = 0.1, 0.3, 1, and $10~\msunyr$,
starting with a $10~\msun$ initial model which has an inflated
structure. We first checked that our calculations well reproduce
the previous results by \citet{Hosokawa12,Hosokawa13},
who study the evolution before the GR effect becomes important.
We terminate the calculations when the stellar central temperature
reaches $\log T_{\rm c} ({\rm K})= 9.2$,
after which the evolutionary timescale is so short 
that the stellar mass hardly increases thereafter. 
The final mass and thus the final fate of the star 
are determined at this epoch.
  

\section{Results} \label{sec:Results}

\begin{figure}[!t]
\figurenum{1}
\includegraphics[width=8cm]{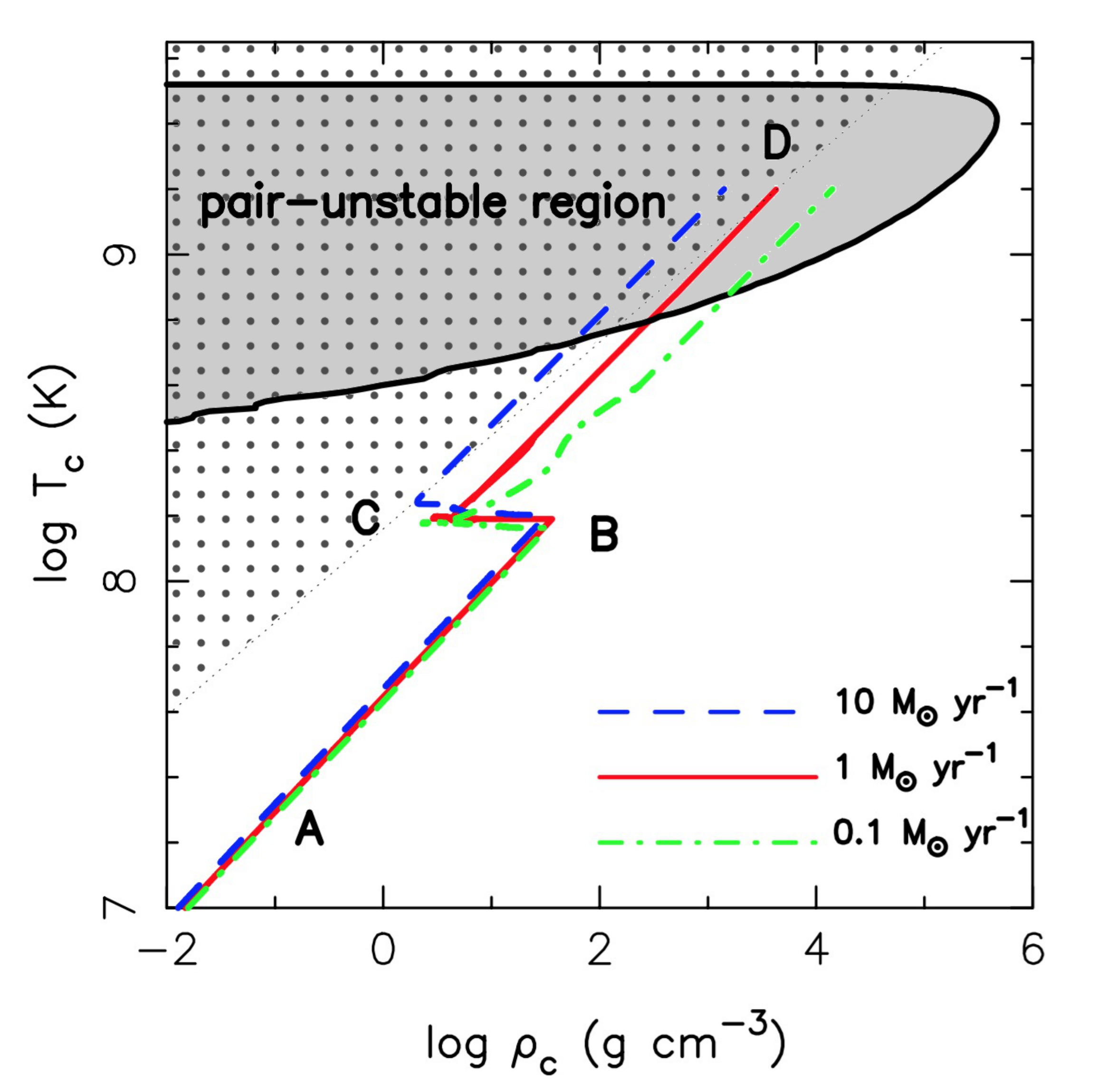}
\caption{The evolution of the central temperature and density for 
the cases with different accretion rates of
$\dot{M} = 0.1$, 1, and $10~\msunyr$. The gray-shaded area denotes
the electron-positron pair-unstable region. The star becomes
GR-unstable in the dot-shaded area
under the assumptions of the $n=3$ polytropic structure and primordial composition.
}
\label{fig:f1}
\end{figure}

\begin{figure}[ht!]
\figurenum{2}
\includegraphics[angle=-90, width=8cm]{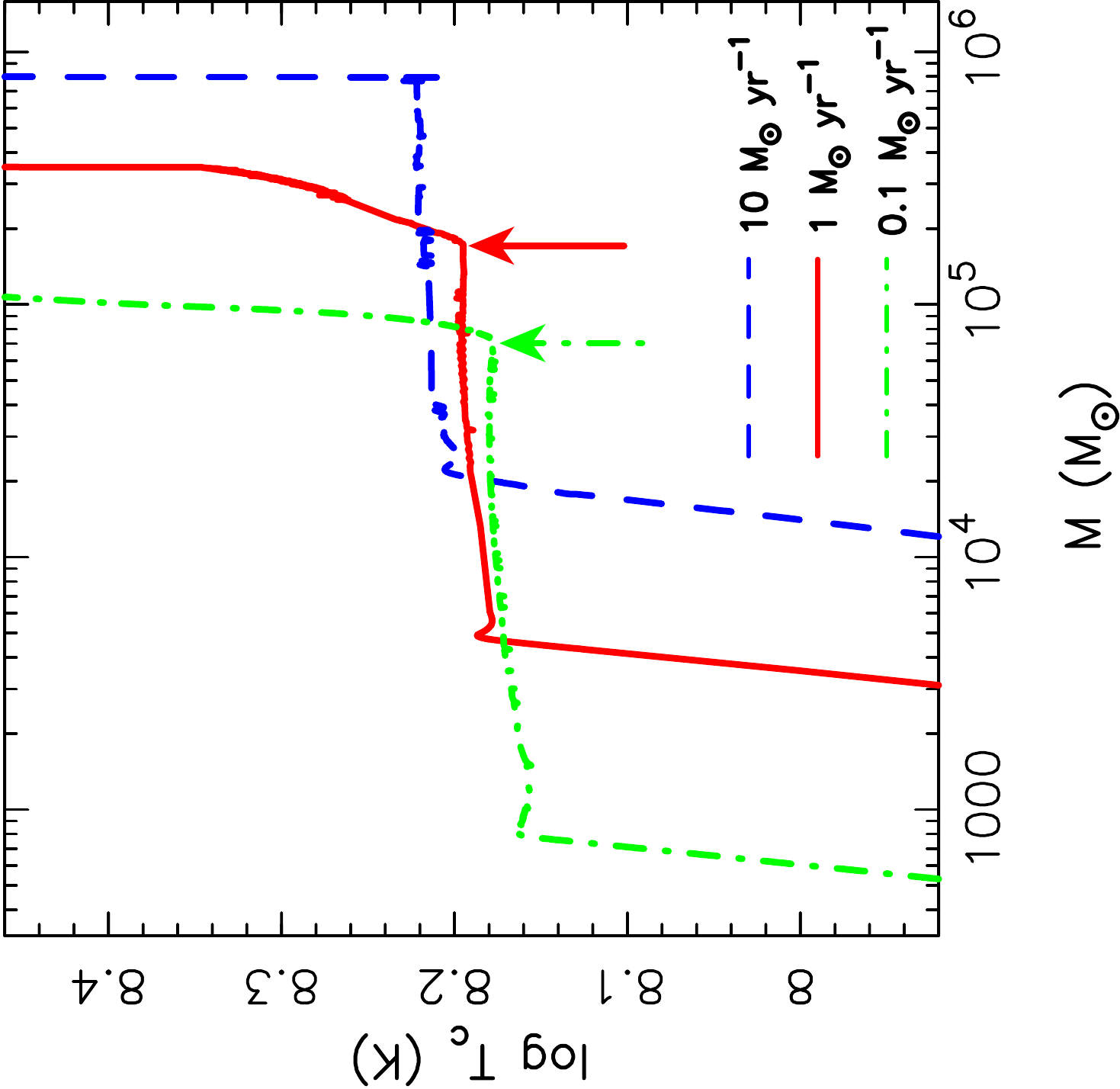}
\caption{
The evolution of the central temperature $T_{\rm c}$ 
as the stellar mass increases. 
Only the values around the stable nuclear burning are shown for clarity.
The different lines represent the different accretion rates,
$10~\msunyr$ (blue dashed), $1~\msunyr$ (red solid),
and $\dot M = 0.1~\msunyr$ (green dot-dashed line).
The vertical arrows indicate the points where the central hydrogen
is exhausted with $\dot{M} = 0.1$ and $1~\msunyr$. }
\label{fig:f2}
\end{figure}

 In Figure~\ref{fig:f1} we show the evolution of the central temperature 
and density for the cases with $\dot M $= 0.1, 1 and $10~\msunyr$. 
Although the evolutionary tracks in this figure are quite similar for all the models,
there are some differences in the actual evolution, 
including the cause of the final collapse, as described below.
From A to B in Figure~\ref{fig:f1}, the stellar core contracts monotonically
over the Kelvin-Helmholtz timescales until the hydrogen burning 
via the CNO cycle starts to sustain the star.
Although the evolutions from A to B are almost identical with different 
$\dot M$ values in Figure~\ref{fig:f1}, Figure~\ref{fig:f2} shows that the 
stellar mass at point B is larger for higher $\dot M$. 
As shown in these figures, the central temperature 
at point B, i.e., when the hydrogen burning begins, 
is $\log T_{\rm c} (\rm K) \simeq 8.2 $ with only very weak dependence on $\dot M$. 
During the hydrogen burning, i.e., from B to C in Figure~\ref{fig:f1}, 
the central temperature remains almost constant while
the density decreases by more than a factor of 10. 
This is because the central entropy is in general higher
for more massive stars, and thus an accreting star evolves 
toward higher entropy (lower density) regions as its mass increases.
After either the exhaustion of nuclear fuel at the center 
(for the cases with 0.1 $\msunyr$) or the onset of the GR instability 
(for 1 and 10 $\msunyr$), the stellar core contracts again toward D.
Figure~\ref{fig:f2} shows that this occurs at higher stellar mass
for higher accretion rate.
In all the models, the stars eventually enter the electron-positron pair-unstable region. 
It is known that stars with $M \simeq 300 - 1000~\msun$ 
collapse directly to form BHs without explosion 
once entering this region \citep[e.g.,][]{Takahashi14}.
Presumably, our SMSs also will follow the same fate 
(see Section~\ref{sec:Discuss}).


\begin{figure*}[t!]
\figurenum{3}
\includegraphics[angle=-90, width=18cm]{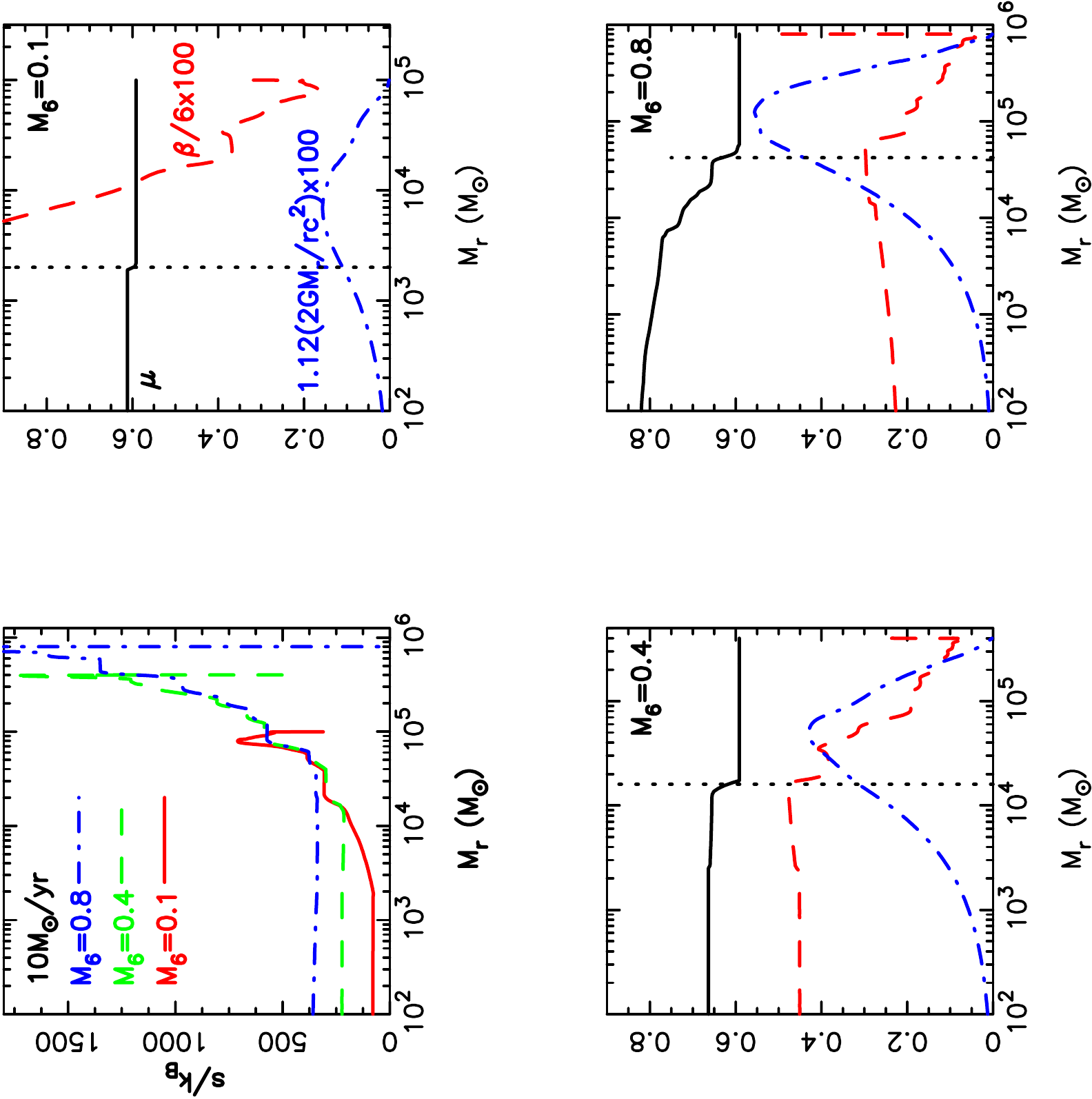}
\caption{
The evolution of the stellar interior structure 
with the accretion rate $\dot{M} = 10~\msunyr$.
The upper left panel shows the evolution of the entropy distribution 
inside the star. The different lines represent
the different epochs when the stellar mass is $10^5~\msun$ (red solid),
$4 \times 10^5~\msun$ (green dashed), and $8 \times 10^5~\msun$ 
(blue dot-solid line). 
The other panels show the development of the GR instability at
the same epochs, at $M = 10^5~\msun$ (upper right), 
$4 \times 10^5~\msun$ (lower left), and $8 \times 10^5~\msun$ (lower right).
In each panel, the radial distributions of $\beta/6$, 
$1.12 (2 G M_r / r c^2)$, and $\mu$ are presented with the
red dashed, blue dot-solid, and black solid lines, respectively. 
The former two quantities are multiplied by a factor of 100.
The vertical dotted lines indicate the mass 
coordinates corresponding to the surface of the isentropic core
(also see text).}
\label{fig:f3}
\end{figure*}

Table~\ref{tb:t1} summarizes the stellar masses 
and mass fractions $X$ or $Y$
for the different accretion rates 
when the central temperature reaches $\log T_{\rm c} (\rm K)= 8.7$.
We here show the mass fractions not at the very center
but at the slightly outer part $M_r = 0.1~M_{\rm core}$, where 
$M_{\rm core}$ is the core mass, 
because the central composition changes relatively fast 
in the final collapse stage.
The central hydrogen has been exhausted
by the onset of the collapse in the cases with $\dot M \leq 1 \msunyr$, for which
the values of the helium fraction $Y$ are described in Table~\ref{tb:t1}.
On the other hand, we show the hydrogen mass fraction $X$ for the model
with the highest rate $\dot M = 10 \msunyr$, where the hydrogen 
still remains at the center.

The GR instability of SMSs has been studied for stellar models with 
the polytropic equation of state with $n=3$ assuming that 
the entire star is radiation-pressure dominant.
For the $n=3$ polytropic star, the critical central temperature above which 
the GR instability occurs can be written as
\citep{ST83},
\begin{equation}
 T_{\rm crit} \simeq 2.5 \times 10^7 (0.5/\mu) M_6^{-1} (\rm{K}),
\end{equation}
where $M_6 \equiv M / 10^6~\msun$ and $\mu$ is the mean molecular weight.
As shown in Figure~\ref{fig:f2}, the central temperature 
during the stable hydrogen and helium burning phase
is roughly constant at $\log T_{\rm c} (\rm K) \simeq 8.2$ for all the models. 
For this temperature, the critical mass for the GR instability is given by
\begin{equation}
 M_6^{\rm crit} \simeq 0.16 \times (0.5/\mu), 
\label{eq:mcr}
\end{equation}
where $M_6^{\rm crit}$ is normalized with $10^6 M_\odot$.
With the mean molecular weight $\mu \simeq (2X+3Y/4)^{-1}= 0.59$ 
for the initial primordial abundance (1.33 for the pure helium matter),  
the critical mass is $M_6^{\rm crit} \simeq 0.14$ (0.06, respectively).


As shown in Table~\ref{tb:t1}, the final mass in our case with $\dot M = 0.1~\msunyr$ 
is $M_6 = 0.12$, smaller than the critical mass estimated above. 
In fact, the collapse in this case is not triggered by the GR instability.
The star just collapses after exhausting the central hydrogen and helium.
For $\dot M \gtrsim 0.3~\msunyr$, on the other hand, the collapse occurs 
while the nuclear fusion is still in operation, suggesting that the 
GR instability causes the collapse. 
The final masses in these cases are, however, again much larger than
the critical masses for the polytropic stars given by equation (\ref{eq:mcr}).


To understand these differences, we show in Figure~\ref{fig:f3} 
(top left) the entropy distribution inside the star at three different epochs
for the case with $\dot M = 10~\msunyr$, 
where effects of the GR instability is the most prominent.
Recall that the polytrope models presume the spatially constant
specific entropy in the interior \citep[e.g.,][]{Kip12}.
As seen in Figure~\ref{fig:f3}, however, 
this is not the case for our accreting stellar models, 
for which the outer envelope has higher entropy than in the core.
Since such an envelope is gravitationally loosely bound, 
higher total stellar mass is required for the GR instability 
in the accreting models than in the polytropic model.  


Another useful expression for the GR-instability criterion is given by \citet{Chandra64}
for a polytropic star with the index $\Gamma (\equiv 1+1/n)$, which states that 
the star is GR unstable if $\Gamma < 4/3 + 1.12~(2 G M/R c^2)$,
where $R$ is the stellar radius.
For a highly radiation-dominant state as in SMSs,
the adiabatic index in the equation of state can be approximated as 
$\Gamma = 4/3 + \beta/6 + O(\beta^2)$, where $\beta (\ll 1)$ 
is the ratio of the gas pressure to total pressure. 
The criterion for the GR instability can thus be written as 
\begin{equation}
 \beta/6 < 1.12~(2 G M/R c^2). 
\label{eq:gr}
\end{equation}
Note that this condition is obtained for a spatially constant $\Gamma$, 
i.e., for a constant $\beta$, 
which is inversely proportional to the specific entropy $s$ as 
$\beta \simeq 4k_{\rm B}/\mu s$ for $\beta \ll 1$. 
In our case, this condition should be applied to the isentropic core.


In Figure~\ref{fig:f3} we show the distributions of $\beta/6$, 
$1.12~(2 G M_r/r c^2)$, and $\mu$ for $\dot M = 10~\msunyr$ 
at three different epochs with $M_6 = 0.1, 0.4$ and 0.8.
We define the surface of the isentropic core (with mass $M_{\rm s}$) 
as the shell where the specific entropy becomes larger than the central value by 1\%;
$M_{\rm s, 6} =  0.002, 0.016$ and $0.042$ for 
$M_6 = 0.1, 0.4 $ and 0.8, respectively.
As seen in the figure, as the stellar mass $M_6$ increases,
$\beta$ becomes smaller while $M_r/r$ becomes larger at the same mass shell $M_r$. 
Owing to the high value of the specific entropy outside the core, 
$\beta$ becomes small in this region with the relation $\beta \simeq 4k_{\rm B}/\mu s$. 
The mean molecular weight $\mu$ increases toward the center as the hydrogen burning proceeds. 
The rise of $M_r/r$ with the total stellar mass indicates that 
the more massive star is more concentrated, as in the case for the ordinary 
primordial stars \citep[e.g.,][]{Marigo01,Marigo03}.
Figure~\ref{fig:f3} shows that, for $M_6 = 0.1$, 
the condition (\ref{eq:gr}) is not satisfied anywhere and the star is GR-stable. 
For $M_6 = 0.4$, the isentropic core is still GR-stable as
condition (\ref{eq:gr}) holds for $M_{r,6} \geq 0.04$, 
only outside the isentropic core $M_{s,6} = 0.016$.
Finally, when the stellar mass reaches $M_6 = 0.8$, 
equation  (\ref{eq:gr}) is satisfied at the surface 
where $M_r = M_s$, and the core begins to collapse.
Similarly, we find that the GR instability begins to operate satisfying
the condition (\ref{eq:gr}) at some moment 
for the cases with $\dot M = 0.3$ and $1.0~\msunyr$, but never with $\dot M = 0.1~\msunyr$.
From this analysis, we conclude that the GR instability causes the 
gravitational collapse for the models with $\dot M \gtrsim 0.3~\msunyr$.

\section{Discussions and Implications} \label{sec:Discuss}

We first compare our results with the previous work by
\citet{Fuller86}, who study the GR stability of hydrogen-burning
stars with fixed masses. 
They define a star $stable$ if the GR instability does not occur 
during the hydrogen burning stage. 
The critical stellar mass for the stability $M_{\rm crit}$
lies between $M_6$ = 0.1 and 0.25. 
However, our calculations show that
all the models except for $\dot M=10~\msunyr$ are ``stable'' and 
begin collapsing after the exhaustion of the hydrogen at the center, 
when the stellar mass is well above $M_{\rm crit}$.
Specifically, the star becomes GR-unstable during the helium-burning stage 
for $\dot M= 0.3 \sim 1.0~\msunyr$ with the final mass $M_6 = 0.19 \sim 0.35$. 
The main difference from \citet{Fuller86} is that 
accreting stars that we consider here have a high-entropy envelope.
Hence the final masses (including the outer envelope) are larger than $ M_{\rm crit}$.


The final stellar masses obtained in this {\it Letter} should be regarded 
as upper mass limits for the remnant BHs after the collapse.
\citet{Chen14} argue that energetic supernova explosions occur at the collapse of 
SMSs in a very narrow mass strip around $M = 55000~\msun$.  
Since none of our models end up in this mass range, it is unlikely 
that such explosions are the final fate of our models.
We will provide further results for this issue using a 1D-GR 
hydrodynamical code in a forthcoming paper.
Preliminary results suggest that the BHs with approximately the same mass 
as the progenitors will be left behind after the collapse.


If the progenitor star is a rapid rotator, some mass ejection or even 
explosion may occur at the final stellar collapse.
An accretion disk can appear around a rotating BH 
and some outflows or jets can be launched.
\citet{Shibata16}, for instance, show by way of a 
GR hydrodynamics simulation that
some portion of the stellar envelope is ejected 
while most of the stellar matter is drawn into a newborn BH.


We conclude that an accreting protostar at a very large
rate of $\dot{M} = 0.1-10~\msunyr$ grows to become
a SMS with mass $1.2-8.0 \times 10^5~\msun$. 
BHs with roughly the same mass will be left as remnants of such SMSs,
which might have sown seeds for the formation of supermassive BHs 
in the early universe.


{\acknowledgements
 We thank Masaru Shibata and Koh Takahashi for valuable comments. 
This work has been supported by Grants-in-Aid for Scientific Research
(26400220: HU, 16H05996, 15H00776, 25800102: TH, 25287040: KO, 25287050: NY) from Japan Society for 
the Promotion of Science and Grant-in-Aid for Scientific Research on Innovative Areas 
(26104007: HU) from the MEXT in Japan.
}

\end{document}